\documentclass[aps,prb,twocolumn, showpacs]{revtex4}

\usepackage{epsfig}
\usepackage{graphicx}

\def\be{\begin{equation}}
\def\ee{\end{equation}}
\def\bea{\begin{eqnarray}}
\def\eea{\end{eqnarray}}

\begin{document}

\title{Exact CNOT Gates with a Single Nonlocal Rotation for Quantum-Dot Qubits}
\author{Arijeet Pal, Emmanuel I. Rashba, and Bertrand I. Halperin}
\affiliation{Department of Physics, Harvard University, Cambridge, Massachusetts 02138, USA}  
\date{\today}
\begin{abstract}
We investigate capacitively coupled two-qubit quantum gates based on quantum dots. For exchange-only coded qubits electron spin $S$ and its projection $S_z$ are exact quantum numbers. Capacitive coupling between qubits, as distinct from interqubit exchange, preserves these quantum numbers. We prove, both analytically and numerically, that conservation of the spins of individual qubits has dramatic effect on performance of two-qubit gates. By varying the  level splittings of individual qubits, $J_a$ and $J_b$, and the  interqubit coupling time $t$,  we can find an infinite number of triples $(J_a, J_b, t)$ for which the two-qubit entanglement, in combination with appropriate single-qubit rotations, can produce an exact CNOT gate. 
This statement is true for practically arbitrary magnitude and form of capacitive interqubit coupling. Our findings promise a large decrease in the number of nonlocal (two-qubit) operations in quantum circuits.
\end{abstract}


\maketitle

\narrowtext

\section{Introduction}
\label{sec:intro}

Quantum computing is envisioned as a sequence of two-qubit operations, which are nonlocal and produce entanglement,  accompanied by a set of proper single-qubit (local) rotations. Two-qubit operations are at the heart of the process and are generally the most demanding.  The ``controlled-not gate" (CNOT)  is a paradigmatic universal nonlocal gate, which allows one to perform, in conjunction with single-qubit rotations, all multi-qubit operations. The controlled-Z gate differs from CNOT only by two Hadamard gates, which are local, so our considerations below are applicable to both gates.\cite{NC}

Progress achieved in developing high-fidelity single qubits based on quantum dots makes two-qubit (and multi-qubit) operation of entangled qubits a timely goal.\cite{Loss,Kloeffel} Some achievements in two-qubit entanglement have been already reported.\cite{Weperen,Shulman,Nowack2,Veldhorst}

Typically, quantum-dot spin qubits consist of one,\cite{Loss} two,\cite{Levy} or three\cite{DiVincenzo} strongly coupled dots, and they differ in the mechanism through which spin rotations are achieved.\cite{RMP07} In particular, they can be driven by {\it ac} magnetic field\cite{Koppens,Veldhorst2} or by {\it ac} electric field through interactions mediated by non-equilibrium nuclear magnetization,\cite{Science05,LairdSST,Bluhm,Maune}  inhomogeneous magnetic fields produced by micromagnets,\cite{Pioro08,Yoneda,Kawakami} or spin-orbit coupling.\cite{Berg} A special class of qubits proposed by DiVincenzo {\it et al.}\cite{DiVincenzo} are exchange-only three-electron qubits, where the qubit is encoded in the subspace with 
total electron spin $S=1/2$ and  projection $S_z=1/2$.   Such qubits, which  were first realized by Laird {\it et al.}\cite{Laird},  exist in three-dot\cite{Sachrajda,Jim1,Jim2, Eng} and two-dot\cite{Shi,Kim} modifications. 
At a proper detuning,\cite{Jim2} the energy spectrum of these systems includes a pair of close levels forming a qubit that can be described in terms of an effective spin, and we use this description in what follows. Single qubit operations, including transitions between the levels, is performed by gate voltages which couple to the  electric charges on  the dots, and take advantage of the tunneling matrix elements that allow charge exchange between the dots.  In the absence of spin-orbit coupling, magnetic field gradients or hyperfine coupling to nuclei, the total electron spin $S$ and its projection $S_z$ are good quantum numbers, which are conserved by the single qubit operations.

Entanglement between two quantum-dot qubits can be established through capacitive and/or exchange coupling. As applied to coded qubits, such suggestions have been made in Refs.~[\onlinecite{Taylor,PRH}] and [\onlinecite{Fong,Doherty,Setiawan}], respectively.  Capacitive coupling, which depends only on the  Coulomb interactions between the qubits, does not violate conservation  of $S$ and  $S_z$ for each  qubit separately, as long as higher energy states can be ignored, this allows both qubits to remain in the coding subspace.  Exchange coupling has the advantage that it is usually stronger, and easier to turn on and off, which should allow faster performance of individual two-qubit operations.  However, exchange interactions do not conserve $S$ and $S_Z$ for the individual qubits, so that in general, after an interval of exchange coupling, the qubits will no longer be in the coding subspace. In order to rectify this, a  proper two-qubit operation requires a sequence of many exchange couplings, interspersed with single qubit operations,  
\cite{Doherty,Setiawan,Mehl}   
and  the total number of operations is very large. Even with high speed and accuracy of individual operations,\cite{Sachrajda,Jim1,Jim2,Kim} the cost of a long chain of them could be rather high.  Thus, there may be an overall advantage to using capacitive coupling rather than exchange coupling for two-qubit operations.
However, one may ask whether capacitive coupling allows one to perform the CNOT operation exactly and, if so, how many nonlocal rotations are required for the  operation.

In this paper we concentrate on three-dot three-electron qubits\cite{Laird,Jim1,Jim2,PRH} with exchange electron-electron interaction within the qubits and capacitive coupling between them.  We prove that CNOT can be performed {\it exactly} using  a {\it single nonlocal operation} (accompanied by local rotations) by proper choice of three parameters, two single-qubit level splittings $(J_a,J_b)$ and the interqubit coupling time $t$, where the subscripts $a$ and $b$ label the two qubits. In fact, there is a countable (infinite discrete) set of such $(J_a,J_b,t)$ triplets. (We assume, here, that during a nonlocal rotation, the capacitive coupling between the qubits, described by four parameters $g_{zz} , g_{xx}, g_{xz},  g_{zx}$  which  depend on the geometry, can be turned on and off sharply, but that the ratio between them is fixed. See the definitions in Section~\ref{sec:hamiltonian}, below.) 
For large ratios of the intra- to inter-qubit couplings, the $(J_a,J_b)$ pairs are arranged on a square lattice, with both  $(J_a,J_b)$ either even or odd (in properly chosen units), but the precise positions of the solutions vary continuously as one varies the relative values of the coupling parameters    $g_{zz} , g_{xx}, g_{xz},  g_{zx}$.

The existence of an extensive set of {\it exact} CNOT gates in a  capacitively-coupled qubit array is the {\it central result} of our paper. While it has been proven for a specific model of qubits, we expect that it is more generic.  Indeed, it has been shown, in recent work by Calderon-Vargas and Kestner,\cite{calderon} that similar effects can be obtained for  capacitaively coupled two-electron singlet-tri[let qubits. These results call  for developing techniques for the control of capacitive couplings and turning them on and off.

The general outline of the paper is as follows. In Sec.~\ref{sec:hamiltonian}, we present a generic four-parameter form of the Hamiltonian of two capacitively coupled coded qubits and choose the basic model to which most of the numerical work is related. In Sec.~\ref{sec:makhlin}, we discuss the transformation from the standard to magic basis and outline our approach based on calculation of Makhlin invariants.\cite{Makhlin} In Sec.~\ref{sec:bert}, we provide an analytic solution for a model with two coupling constants, $(g_{zz},g_{xx})$,  which unveils the basic potentialities of two capacitively coupled qubits, including the existence of an infinite set of exact CNOT gates. In Sec.~\ref{sec:large}, we find the asymptotic behavior of invariants in the limit where  the intraqubit exchange splitting is large compared to the interqubit coupling limit, for a generic-form interqubit coupling, and we derive the asymptotic map of $(J_a,J_b,t)$-sets producing CNOT gates. Some general properties of the nonlocal part $M_d$ of the evolutionary matrix $M_S$ are investigated in Sec.~\ref{sec:KC}. In Sec.~\ref{sec:CNOT}, we build a map of the positions of CNOT gates in $(J_a,J_b)$ plane for a generalized basic model as a function of a scaling parameter, and we investigate analytical properties of both Makhlin invariants near the CNOT points. Sec.~\ref{sec:Mdm} includes matrix identities specific for CNOT gates, and Sec.~\ref{sec:LR} discusses in brief the local rotations that should be performed for achieving Controlled-Z gates. Sec.~\ref{sec:summary} summarizes our results.

\section{Hamiltonian}
\label{sec:hamiltonian}

In this paper we consider three-electron three-dot coded qubits\cite{DiVincenzo} in which two-axis rotations are performed by Heisenberg exchange only, without involvement of any additional mechanisms such as spin-orbit coupling, inhomogeneous magnetic field, or nonequilibrium nuclear polarization. Efficient operation of such qubits has been already demonstrated experimentally.\cite{Laird,Sachrajda,Jim1,Jim2}  Three-electron two-dot qubits\cite{Shi,Kim} have very similar properties. 

Depending on the specific shape of two capacitively coupled three-dot qubits and their mutual position, the basic two-qubit configurations may be termed as linear, butterfly, two-corner, and loop geometries.\cite{PRH,Doherty,Setiawan} Details of the geometry mostly determine the interqubit-coupling part $H_{\rm int}$ of the total Hamiltonian $H=H_0+H_{\rm int}$, while the Hamiltonian $H_0$ of individual qubits can always be chosen as
\be
H_0=-\frac{1}{2}(J_a\sigma_z^a+J_b\sigma_z^b).
\label{eq1}
\ee
Here and below, $(\sigma_j^a,\sigma_j^b)$, $j=(x,y,z)$, are Pauli matrices acting on $(a,b)$ qubits, respectively.\cite{PRH} 

\vspace{3mm}
\begin{figure}[!hbtp]
\includegraphics[width=2.5in]{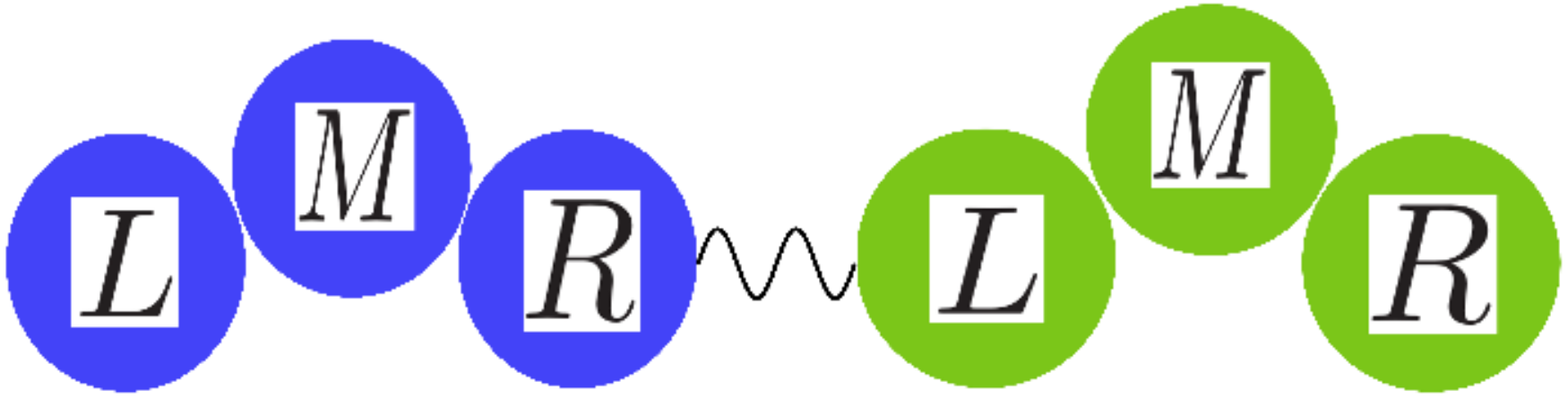}
\caption{(Color online) Model of a two-qubit gate consisting of two three-dot qubits with capacitive coupling between the right dot of qubit $a$ and left dot of qubit $b$.}
\label{fig:geometry}
\end{figure}

Capacitive coupling between qubits can be expressed in terms of a Coulomb interaction between electric charges on pairs of dots belonging to different qubits. Generically, for two-level systems the operators of charges can only include real $2\times2$ matrices, hence, they are linear combinations of two Pauli matrices, $\sigma_x$ and $\sigma_z$, and a unit matrix $\sigma_0$. Explicit expressions for charges were derived in Ref.~\onlinecite{PRH}. The Hamiltonian $H_{\rm int}$ includes Kronecker products of charges in which matrices $\sigma_0^a$ and $\sigma_0^b$ can be omitted because they only result in an additive constant and corrections to $H_0$ that can be eliminated by local spin rotations and/or renormalization of parameters. Therefore, the most general expression for the Hamiltonian $H_{\rm int}$ of capacitive coupling between two qubits reads 
\be
H_{\rm int}=g_{xx}\sigma_x^a\sigma_x^b+g_{xz}\sigma_x^a\sigma_z^b+g_{zx}\sigma_z^a\sigma_x^b+g_{zz}\sigma_z^a\sigma_z^b.
\label{eq2}
\ee
In this paper, we choose the energy scale such that $g_{zz} = 1/4$, and we assume that the other coefficients $g_{ij}$ are either small or of  the order of unity.

Generic properties of capacitive gates in the weak coupling regime, $|g_{ij}|\ll J_a,J_b$, will be derived analytically in Sec.~\ref{sec:large} for arbitrary ratios between the coupling constants $g_{ij}$. However, the space of these constants is too wide to investigate it all numerically. Therefore, we choose as the {\it basic model} a linear geometry with capacitive coupling between the right dot of qubit $a$ and the left dot of qubit $b$, Fig.~1. In this case\cite{PRH} $g_{zz}=1/4,g_{zx}=\sqrt{3}/4,g_{xz}=-\sqrt{3}/4,g_{xx}=-3/4$, and
\be
H_{\rm int}=\left(-\frac{\sqrt{3}}{2}\sigma_x^a+\frac{1}{2}\sigma_z^a\right)\left(\frac{\sqrt{3}}{2}\sigma_x^b+\frac{1}{2}\sigma_z^b\right).
\label{eq3}
\ee
The operators in the parentheses are proportional to the electrical charges at the ends of the two qubits.\cite{charges} As we have used the interqubit couplings in  $H_{\rm int}$, to establish the energy scale we see that  whenever the intraqubit coupling is essentially stronger than interqubit coupling, we have  $J_a,J_b\gg1$. 

In Sec.~\ref{sec:bert} we solve analytically a model with $g_{xz}=g_{zx}=0$ and arrive at an infinite set of CNOT gates. In Sec.~\ref{sec:CNOT} we consider a generalization of the basic model by scaling its parameters $(g_{zx},g_{xz},g_{xx})$ and arrive numerically at a similar picture that reflects generic properties of capacitive two-qubit gates. We have also  confirmed the generality of these conclusions by numerical studies of various other randomly-selected values of $g_{ij}$.

\section{Evolutionary operator and Makhlin invariants}
\label{sec:makhlin}

We choose the standard basis in the $4\times4$ two-qubit space as $|00\rangle,|01\rangle,|10\rangle,|11\rangle$ with $|0\rangle$ for $|\uparrow\rangle$ and $|1\rangle$ for $|\downarrow\rangle$. Then $H_0$ is diagonal
\be
H_0=\left(\begin{array}{cccc}-\frac{J_a+J_b}{2}&0&0&0\\
0&-\frac{J_a-J_b}{2}&0&0\\
0&0&\frac{J_a-J_b}{2}&0\\
0&0&0&\frac{J_a+J_b}{2}\\
\end{array}\right)
\label{eq4}
\ee
and $H_{\rm int}$ of Eq.~(\ref{eq2}) equals
\be
H_{\rm int}=\left(\begin{array}{cccc}
g_{zz}&g_{zx}&g_{xz}&g_{xx}\\
g_{zx}&-g_{zz}&g_{xx}&-g_{xz}\\
g_{xz}&g_{xx}&-g_{zz}&-g_{zx}\\
g_{xx}&-g_{xz}&-g_{zx}&g_{zz}\\
\end{array}\right).
\label{eq5}
\ee
For the basic model, the sum of the squares of matrix elements in each row (column) of $H_{\rm int}$ equals one, which sets the interqubit interaction as the energy scale. In more generic models, we always choose $g_{zz}=1/4$ and in this way set the scale of energy; with $\hbar=1$, this also sets the scale of time $t$.

It follows from these equations that for large $J_a,J_b$,  the off-diagonal part of the evolutionary operator $M_S(t)=\exp{(-iHt)}$ is small in the parameters $J_a^{-1},J_b^{-1}\ll1$.  (We use the  subscript $S$ to indicate  the standard basis.) In what follows, we use side by side with the numerical study also an analytical approach based on $1/J$ expansion. It is seen from Eq.~(\ref{eq4}) that keeping $J_a\approx 2J_b$ allows avoiding small denominators in the perturbation theory. On the contrary, with $J_a\approx J_b$, perturbation theory becomes inapplicable. 

The conditions for a gate be equivalent  to the universal CNOT gate, up to local rotations,  are expressed most conveniently in the ``magic basis" \cite{HW1997} (of time-inversion symmetric Bell states),  which we choose as\cite{Makhlin}
\bea
|\Phi_1\rangle&=&\frac{1}{\sqrt{2}}(|00\rangle+|11\rangle),\,\,|\Phi_2\rangle=\frac{i}{\sqrt{2}}(|01\rangle+|10\rangle),\nonumber\\
|\Phi_3\rangle&=&\frac{1}{\sqrt{2}}(|01\rangle-|10\rangle),\,\,|\Phi_4\rangle=\frac{i}{\sqrt{2}}(|00\rangle-|11\rangle).
\label{eq6}
\eea
In this basis local (single-qubit) gates are represented by real matrices, and the transformation of $M_S$ into the magic basis is performed by $M_B=Q^\dagger M_SQ$, where\cite{Makhlin}
\be
Q=\frac{1}{\sqrt{2}}\left(\begin{array}{cccc}
1&0&0&i\\
0&i&1&0\\
0&i&-1&0\\
1&0&0&-i
\end{array}\right).
\label{eq7}
\ee
The equivalence of two nonlocal gates, up to local operations, is given by two Makhlin invariants $(G_1,G_2)$, which are defined in terms of traces of a unitary matrix $m=M_B^TM_B$ ($T$ stands for transpose) and of its square:\cite{Makhlin,ZVSW}
\be
G_1=\frac{1}{16}{\rm Tr}^2[m],\,\,G_2=\frac{1}{4}({\rm Tr}^2[m]-{\rm Tr}[m^2]).
\label{eq8}
\ee
Here, we have used the fact  that ${\rm Det}[M^\dagger]=1$, which  is true for the above Hamiltonian $H$. The choice of numerical coefficients in (\ref{eq8}) results in values of $(G_1,G_2)$ of the order of unity for most  typical gates.\cite{Makhlin} $G_1$ is a complex number and $G_2$ is real. For CNOT and Controlled-Z gates these invariants must be $G_1=0,G_2=1$, and we prove below that these equations can be fulfilled for an infinite discrete set of real values of three parameters of the theory, $(J_a,J_b,t)$.

Because CNOT and Controlled-Z gates have the same invariants, they are equivalent up to  local rotations, which may be performed by two Hadamard gates.\cite{NC} However, it follows from Eqs.~(\ref{eq4}) and (\ref{eq5}) that for $J_a,J_b\gg1$ the evolutionary matrix $M_S(t)$ is nearly diagonal and therefore is closer to a Controlled-Z matrix (which is diagonal in the standard basis) than to a CNOT matrix. In the standard basis, Controlled-Z is represented by the matrix 
$\frac{1}{2}\sigma_0^b(\sigma_0^a+\sigma_z^a)+\frac{1}{2}\sigma_z^b(\sigma_0^a-\sigma_z^a)$, while CNOT is given by 
$\frac{1}{2}\sigma_0^b(\sigma_0^a+\sigma_z^a)+\frac{1}{2}\sigma_x^b(\sigma_0^a-\sigma_z^a)$.
For this reason, we will sometimes refer simply to Controlled-Z gates. (See Sec.~\ref{sec:LR} below.)

\section{Exactly soluble model}
\label{sec:bert}

With $g_{zx}=g_{xz}=0$, operator $H$ is block-diagonal, and both Makhlin invariants can be calculated explicitly.\cite{footnote}     It is convenient to introduce functions
\be
f_\pm=\frac{(J_a\pm J_b)^2+4g_{xx}^2\cos{(t \sqrt{(J_a\pm J_b)^2+4g_{xx}^2})}}{(J_a\pm J_b)^2+4g_{xx}^2}.
\label{eqA}
\ee
Then the equation ${\rm Tr}[m]=0$ reduces to
\be
e^{it/2}f_-+e^{-it/2}f_+=0.
\label{eqB}
\ee
If $(J_a-J_b)^2>4g_{xx}^2$, then $f_++f_->0$, and the equation for the real part of the trace reduces to $\cos{(t/2)}=0$. We restrict ourselves with the lowest root of the last equation, $t=\pi$. Then it follows from the equation for the imaginary part of the trace that $f_+=f_-$. Because the expression for $G_2$ is
\[
G_2=2f_{+}f_{-}+\cos{t},
\]
the equation $G_2=1$ reduces to $f_+f_-=1$. The only solutions of these two equations for $f_\pm$ are
\be
\sqrt{(J_a\pm J_b)^2+4g^2_{xx}}=2n_{\pm} ,
\label{eqD}
\ee
where $n_\pm$ are positive integers.   
It is convenient to define two integers, 
  $r=n_++n_-,$ and $s=n_+-n_-$,  
so that 
$n_{\pm} = (r \pm s)/2$, where $r$ and $s$ are integers of the same parity, with the additional constraint that $r > |s|  \geq 0$.  
We then arrive at the following equations for  $(J_a,J_b)$:
\be
(J_a \pm J_b)^2=(r \pm s)^2 -4g^2_{xx}
\label{eqEE}
\ee
These equations have real solutions whenever $2|g_{xx}| \leq r-|s|$.

Thus, for a fixed value of  $g_{xx}$, with $ g_{xz}=g_{zx}=0$, we have found  an infinite set of exact CNOT gates,  indexed by pairs of integers $(r,s)$ of the same parity, restricted only by the criterion $(r-|s|)>2|g_{xx}|$.  The shortest operation time of the gates is precisely $t=\pi$ in each case. In the limit $g_{xx} \to 0$, the solutions for $(J_a, J_b)$ approach the integer values $\pm (r,s)$ or $\pm (s,r)$. As the value of $|g_{xx}|$ is increased, for fixed values  of $(r,s)$, the solutions for $(J_a, J_b)$ move along the hyperbola
\be
J_aJ_b=rs.
\label{eqE}
\ee
As long as $2|g_{xx}|< r-|s|$, there will be two different solutions on each branch of the hyperbola, related to each other by interchange of the values of $J_a$ and $J_b$.  When $2|g_{xx}|\to r-|s|$, the two solutions on each branch come together on the axis $J_a= \pm J_b$ , (depending on the sign of $s$),  and ``annihilate" each other. Real solutions do not exist for $2|g_{xx}|> r-|s|$.

Although exact CNOT operations are possible for both positive and negative values of $J_a$ and $J_b$, as we have seen in this section,  we shall restrict our cosiderations in the remainder of this paper  to the positive quadrant of the $J_a -  J_b$ plane, as this seems to be the region of most immediate physical interest.

\section{Large$-J$ expansion}
\label{sec:large}

Our next insight onto the optimal sets of $(J_a,J_b,t)$ comes from the large-$J$ expansion of the operator $M_S(t)$. To eliminate secular terms in the expansion, the diagonal part of $H_{\rm int}$ with $g_{zz}=1/4$ was included in $H_0$; no restrictions are imposed on other coupling constants $g_{ij}$. We note that while $g_{zz}$ is small compared with $(J_a,J_b)$, it is the only term in $H_0$ resulting in entanglement.

Formal expansion in $1/J\ll1$ is complicated by poles at $J_a=J_b$ as seen from Eqs.~(\ref{eq9}) and (\ref{eq10}) below. To overcome this problem it is convenient to introduce a formal parameter $\gamma$ as $H_{\rm int}\rightarrow \gamma H_{\rm int}$, find expansion in $\gamma$, and finally put $\gamma=1$. Then consecutive  steps include transforming $M_S(t)$ into the interaction representation, calculating it in the second order in $\gamma$, transforming the result into the magic basis, and calculating the $m$-matrix, its trace ${\rm Tr}[m]$, and the  invariants $(G_1,G_2)$. Final results were simplified by using $J_a,J_b\gg1$. For technical reasons it is convenient to use the condition ${\rm Tr}[m]=0$ instead of $G_1=0$. While the expansion of $m$ includes $1/J$ terms, the  expansion of ${\rm Tr}[m]$ consists of a zero-order term followed with $1/J^2$ terms. After tedious calculations, we arrive at a $1/J^2$ order result:
\bea
{\rm Tr[m]}&=&4\cos{\frac{t}{2}}+g_{xx}^2F_{xx}+g_{xz}^2F_{xz}+g_{zx}^2F_{zx},\nonumber\\
F_{xx}&=&-16\bigg\{e^{\frac{it}{2}}\frac{\sin^2[\frac{1}{2}(J_a-J_b)t]}{(J_a-J_b)^2}\nonumber\\
&+&e^{-{\frac{it}{2}}}\frac{\sin^2[\frac{1}{2}(J_a+J_b)t]}{(J_a+J_b)^2}\bigg\},\nonumber\\
F_{xz}&=&\frac{4}{J_a^2}\left[-4\cos{\frac{t}{2}}+4\cos(J_at)+t\sin{\frac{t}{2}}\right],\nonumber\\
F_{zx}&=&\frac{4}{J_b^2}\left[-4\cos{\frac{t}{2}}+4\cos(J_bt)
+t\sin{\frac{t}{2}}\right],\nonumber\\
\label{eq9}
\eea
and
\bea
&&G_2=(2+\cos{t})-16g_{xx}^2\frac{J_a^2+J_b^2}{(J_a^2-J_b^2)^2}\nonumber\\
&\times&[1-\cos{J_at}\cos{J_bt}-\frac{2J_aJ_b}{J_a^2+J_b^2}\sin{J_at}\sin{J_bt}]\nonumber\\
&+&2\left(\frac{g_{xz}^2}{J_a^2}+\frac{g_{zx}^2}{J_b^2}\right)(t\sin{t}-8\cos^2{\frac{t}{2}})\nonumber\\
&+&16\left(\frac{g_{xz}^2}{J_a^2}\cos{J_at}+\frac{g_{zx}^2}{J_b^2}\cos{J_bt}\right)\cos{\frac{t}{2}}.\nonumber\\
\label{eq10}
\eea
The leading cosine terms of both equations originate from the diagonal part of the interaction, $\frac{1}{4}\sigma_z^a\sigma_z^b$.

It is seen from Eqs.~(\ref{eq9}) and (\ref{eq10}) that zero-order solutions of equations ${\rm Tr}[m]=0$ and $G_2=1$ are $t_n=n\pi$, $n$ being integers. Below we restrict ourselves to the shortest operational time, $n=1$. Considering ${\rm Tr}[m]=0$ as an equation for $t$, $t=\pi+\delta t$, we arrive at $\delta t\sim1/J^2$. Then, restricting ourselves to terms of the order of $1/J^2$,  we estimate $2+\cos t\approx1$, the last two lines of Eq.~(\ref{eq10}) can be omitted, and equation $G_2=1$ reduces to
\be
\cos{J_a\pi}\cos{J_b\pi}+\frac{2J_aJ_b}{J_a^2+J_b^2}\sin{J_a\pi}\sin{J_b\pi}=1.
\label{eq11}
\ee
Remarkably, the equation does not depend on the coupling constants $g_{ij}$. 
Expressing the trigonometric functions in (\ref{eq11}) in terms of $\tan{(J_a\pi/2)}$ and $\tan{(J_b\pi/2)}$, one can check that the only {\it real} solutions of Eq.~(\ref{eq11}) are $J_a=r, J_b=s$, $r$ and $s$ being {\it integers of the same parity}.  

Because the parities of $J_a$ and $J_b$ coincide, $\cos{J_a\pi}=\cos{J_b\pi}\equiv\cos{J\pi}$, and the condition ${\rm Tr}[m]=0$ allows estimating $\delta t$ as
\be
\delta t\approx2(g_{xz}^2+g_{zx}^2)(J_a^{-2}+J_b^{-2})(4\cos{J\pi}+\pi).
\label{eq12}
\ee
Remarkably, due to the last factor the values of $\delta t$ have opposite signs for odd and even values of $J$, and positive values are larger than negative ones by an order of magnitude.

The above results suggest existence of an infinite two-parameter set of discrete solutions of the equations for CNOT gates, $G_1=0$ and $G_2=1$, which in the large $J$ limit are $t\approx\pi$, with $(J_a,J_b)$ being integers of the same parity; notice excellent qualitative agreement with the results of Sec.~\ref{sec:bert}. Because the results of this section are valid only with $1/J^2$ precision, it still should be checked whether equations $G_1=0,G_2=1$ have exact solutions or are satisfied only approximately, in some order in $1/J$. We prove numerically in Sec.~\ref{sec:CNOT} below the existence of exact solutions obeying the basic patterns described above, and this provides extension of the analytical results of Sec.~\ref{sec:bert} to generic interqubit coupling. Hence, two capacitively coupled qubits can serve universally as perfect CNOT (Controlled-Z) gates.

In the numerical work below we will be looking for solutions of equations $G_1=0,G_2=1$ as zeros of the function 
\be
G=\sqrt{|G_1|^2+(G_2-1)^2}.
\label{eq13}
\ee
In this connection it is worth mentioning that the expression in second line of Eq.~(\ref{eq10}) can be rewritten as
\bea
&&\frac{J_a^2+J_b^2}{(J_a-J_b)^2}[1-\cos{J_at}\cos{J_bt}-\frac{2J_aJ_b}{J_a^2+J_b^2}\sin{J_at}\sin{J_bt}]\nonumber\\
&=&1-\cos{J_at}\cos{J_bt}+\frac{4J_aJ_b}{(J_a-J_b)^2}\sin^2{\left[(J_a-J_b)\frac{t}{2}\right]}.
\label{eq14}
\eea
For $J_a\approx2J_b$, the coefficient $4J_aJ_b/(J_a-J_b)^2\approx8$ and generically the last term dominates in second line of (\ref{eq14}). However, at $t=\pi$ it vanishes along the lines $J_a-J_b=2m$, $m$ being integers. Because this term also makes a significant contribution to $G$, deep valleys in the plots of $G(J_a,J_b,t)$ along the $(J_a-J_b)=2m$ lines are expected whenever $t\approx\pi$.  (Cf. Fig.~4 below.) 

\section{Nonlocal part of $M$-matrix}
\label{sec:KC}

Before switching to numerical results, it is instructive to take a closer look into some exact results related to invariants $(G_1,G_2)$. According to Kraus and Cirac\cite{KC2001} (see also Ref.~\onlinecite{Khaneja}), the  matrix $M_S$ can be represented as
\be
M_S=U_{ab}M_dV_{ab},
\label{eq15}
\ee
where $U_{ab}$ and $V_{ab}$ are Kronecker products of local unitaries, $U_{ab}=U_a\otimes U_b$ and $V_{ab}=V_a\otimes V_b$, and $M_d=\exp{(-iH_d)}$ is a unitary operator creating entanglement of two qubits. Here
\be
H_d=\alpha_x\sigma_x^a\sigma_x^b+\alpha_y(\sigma_y^a)^T\sigma_y^b+\alpha_z\sigma_z^a\sigma_z^b,
\label{eq16}
\ee
where the coefficients $(\alpha_x,\alpha_y,\alpha_z)$ are real numbers. One can check by inspection that
\bea
Q^\dagger\sigma_x^a\sigma_x^bQ&=&\sigma_z^a\sigma_0^b,\nonumber\\
Q^\dagger\sigma_y^a\sigma_y^bQ&=&\sigma_0^a\sigma_z^b,\nonumber\\
Q^\dagger\sigma_z^a\sigma_z^bQ&=&\sigma_z^a\sigma_z^b,
\label{eq17}
\eea
where $(\sigma_0^a,\sigma_0^b)$ are unit matrices in the corresponding $2\times2$ subspaces. Therefore, $M_d$ is diagonal in the magic basis. Because products of local unitaries $U_{ab}$ and $V_{ab}$ do not change entanglement, the invariants $(G_1,G_2)$ of matrix $m$ are equal to the invariants $(G_1^d,G_2^d)$ of matrix $m_d=(M_d^B)^TM_d^B$ with properly chosen coefficients $(\alpha_x,\alpha_y,\alpha_z)$. An easy calculation results in
\bea
{\rm Re}[G_1^d]&=&\cos^2{2\alpha_x}\cos^2{2\alpha_y}\cos^2{2\alpha_z}\nonumber\\
&-&\sin^2{2\alpha_x}\sin^2{2\alpha_y}\sin^2{2\alpha_z},\nonumber\\
{\rm Im}[G_1^d]
&=&\frac{1}{4}\sin{4\alpha_x}\sin{4\alpha_y}\sin{4\alpha_z},\nonumber\\
G_2^d&=&\cos{4\alpha_x}+\cos{4\alpha_y}+\cos{4\alpha_z}\nonumber\\
&=&4{\rm Re}[G_1^d]-\cos{4\alpha_x}\cos{4\alpha_y}\cos{4\alpha_z}.
\label{eq18}
\eea
Two first equations coincide with the results of Zhang {\it et al.},\cite{ZVSW} and the expessions for $G_2^d$ are equivalent to their result but are presented in a form more convenient for our goals.

Equations (\ref{eq18}) are periodic in all three variables with a period $\pi/2$. Hence, we confine solutions inside the interval $-\pi/4\leq\alpha_x,\alpha_y,\alpha_z\leq\pi/4$. Additional symmetries are circular permutations of all variables, pair permutations $(\alpha_x\rightarrow\alpha_y,\alpha_y\rightarrow\alpha_x)$,  $(\alpha_x\rightarrow-\alpha_y,\alpha_y\rightarrow-\alpha_x)$, and all permutations similar to them. 
 
It is seen from Eqs.~(\ref{eq18}) that whenever equations for a CNOT gate ${\rm Re}[G_1^d]=0$ and $G_2^d=1$ are satisfied, equation ${\rm Im}[G_1^d]=0$ is satisfied automatically. Therefore, we have two equations on three real parameters $(\alpha_x,\alpha_y,\alpha_z)$. Nevertheless, their only solutions are isolated points:
\be
\alpha_x=0,\,\,\alpha_y=0,\,\, \alpha_z=\pi/4,
\label{eq19}
\ee
and the points that can be obtained from this by the above permutation group. These solutions are saddle points both for ${\rm Re}[G_1^d]$ and $G_2^d$. For them the eigenvalues $e^{-i\lambda_j}$ of $M_d$ are twice degenerate and equal to $e^{\pm i\pi/4}$.\cite{FNER} 
Moreover,  because $m_d$ is unitary equivalent to $M_d^2$, its eigenvalues are equal to $e^{\pm
i\pi/2}$.  (This equivalence is easy to check and will be demonstrated explicitly in
Sec. \ref{sec:Mdm}, below.)
The parameters $\lambda_j$ obey the condition $\sum_{j=1}^4\lambda_j=0$ because ${\rm Tr}[\sigma_z^a\sigma_z^b]=0$. The trace of $M_d$ equals ${\rm Tr}[\exp{(-iH_d)}]=2\sqrt{2}$. These results will be used in Sec.~\ref{sec:Mdm}.

The above conclusions about the set of solutions of equations $G_1=0,G_2=1$ for matrix $M_d=\exp{(-iH_d)}$ have important implications. Because the number of free parameters $(J_a,J_b,t)$ of our model is equal to the number of parameters of Hamiltonian $H_d$, we can expect existence of only a discrete set of $(J_a,J_b,t)$ triplets obeying the requirements of CNOT gates. If they do exist, they are expected to satisfy the conditions $t\approx\pi$ with  $J_a,J_b$ being close to integers of the same parity in the region $J_a,J_b\gg1$, according to the results of Sec.~\ref{sec:large}. In the next section, we (i) prove the existence of such a set by numerical means, (ii) investigate the anomaly along the line $J_a\approx J_b$ where the perturbation expansions of Eqs.~(\ref{eq9}) and (\ref{eq10}) diverge, and (iii) make brief comments about the small $J_a,J_b\sim1$ region.

\section{Exact CNOT gates in $(J_a,J_b)$ plane}
\label{sec:CNOT}

In this section we prove numerically the existence of an infinite discrete set of points in $(J_a,J_b,t)$ space where the invariants $(G_1,G_2)$ of Eq.~(\ref{eq8}) satisfy the conditions $G_1=0,G_2=1$ for CNOT gates. 
We have found that errors accumulate tremendously in the  $M_S$ and $m$ matrices when calculations are performed with the standard precision of Mathematica 10.0. Therefore, the results presented below were derived with the WorkingPrecision$\rightarrow$100. With this precision, we found points  of deep minima of $G(J_a,J_b,t)$,  with magnitude typically of $10^{-50}$,  and  we identify them as {\it exact} zeros of $G$. All presented data were found for shortest pulses of interqubit coupling, with $t\approx\pi$.

Calculations for the basic model of Eq.~(\ref{eq3}) resulted in a set of zeros, $G=0$, which are very close to the predictions of Sec.~\ref{sec:large} for large $J_a,J_b\gg1$ and away from the singular line $J_a=J_b$. As expected, the accuracy of large-$J$ expansion is especially high near $J_a=2J_b$ and $J_b=2J_a$ lines, cf. Sec.~\ref{sec:makhlin}. In the vast region where the accuracy of asymptotic expansion remains satisfactory, zeros of $G$ found numerically can be brought into correspondence with the zeros found in Sec.~\ref{sec:large} unambiguously. With decreasing $(J_a,J_b)$ and/or approaching the $J_a=J_b$ line the deviations of $(J_a,J_b)$ from integer values increase, so to establish the genesis of zeros, we followed their evolution as we changed the coupling constants $g_{ij}$ gradually.  In particular, we scaled  them as $g_{zx}=\lambda\sqrt{3}/4, g_{xz}=-\lambda\sqrt{3}/4, g_{xx}=-3\lambda/4$ with $0\leq\lambda\leq1$ while keeping $g_{zz}=1/4$ constant. This allowed us to demonstrate the existence of exact roots of $G$ in the region of the intermediate parameter values  $3\alt J_a,J_b\alt10$ and relate them unambiguously to their origins at  the integers of Sec.~\ref{sec:large}. Results of these calculations are presented in Fig.~2 and discussed below.

The scaled model reduces to the parameters of our basic model when $\lambda=1$.   At $\lambda=0$, however, the  equations $G_1=0,G_2=1$ reduce to $\cos{(t/2)}=0,\cos{t}=-1$. Their first solution is $t=\pi$ with no restrictions on $(J_a,J_b)$. This coincides with the well known result for Ising gates. However, as soon as $\lambda$ deviates from zero, an infinite discrete set of the roots of equation $G=0$ emerges from the $(J_a,J_b)$ continuum. They correspond to all even-even and odd-odd pairs of $(J_a,J_b)$ values with $J_a,J_b\geq3$, but with  the diagonal $J_a=J_b$ excluded.  (See Fig.~2.) [The behavior observed here for $\lambda \to 0$  is  similar to the behavior we found  for $g_{xx} \to 0$ in the exactly solvable model, with $g_{xz}=g_{zx}=0$, in  Sec. (\ref{sec:bert}).]   With increasing $\lambda$, zeros of $G$ adjacent to the $J_a=J_b$ diagonal move mostly in such a way to fill the void around it. Values of $t$ at $\lambda=1$ shown in Fig.~2 are in a reasonable agreement with the estimate of Eq.~(\ref{eq12}). 

\vspace{3mm}

\begin{figure}[!hbtp]
\includegraphics[width=3.0in]{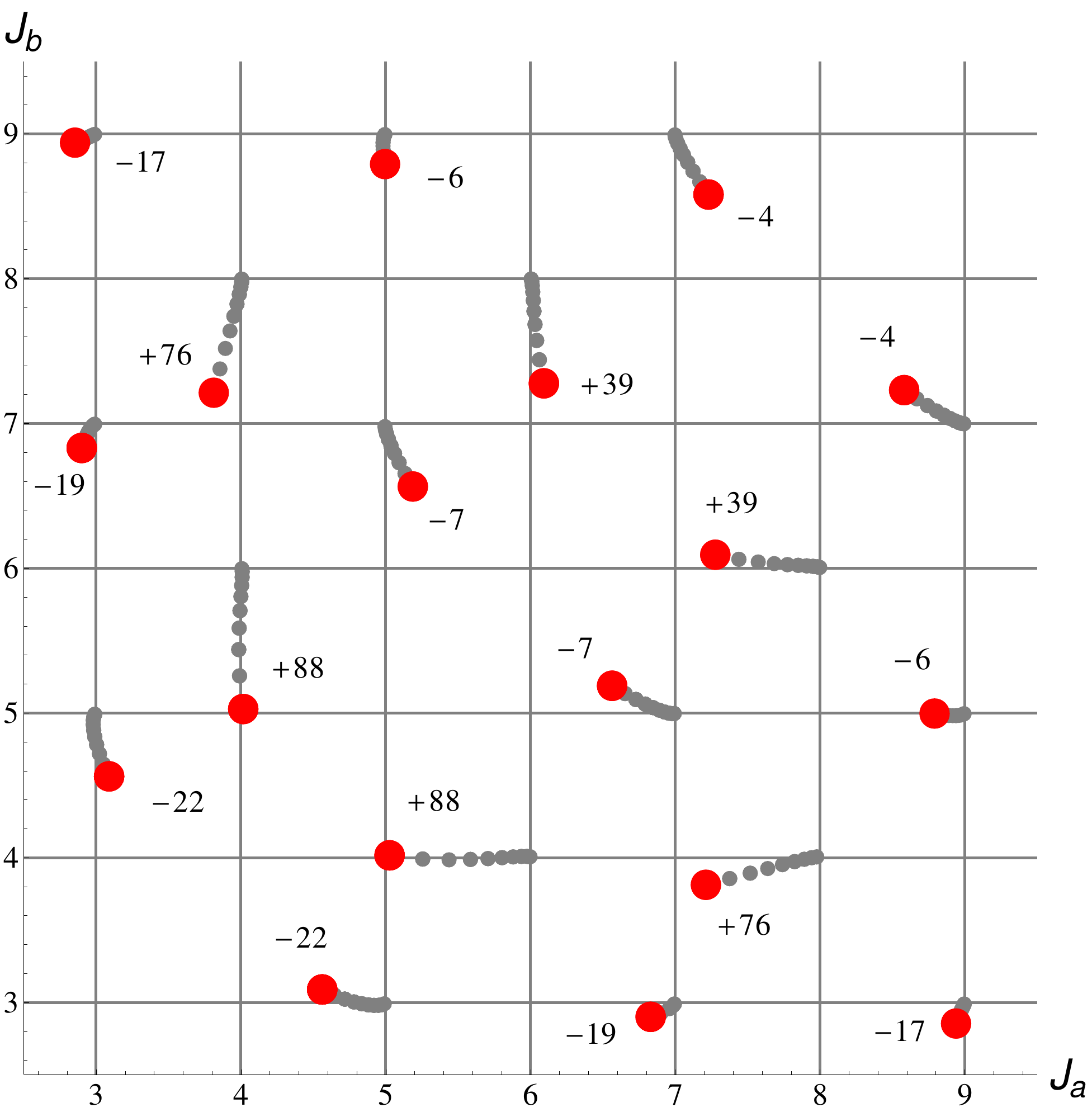}
\caption{(Color online) Trajectories of the roots of equation $G(J_a,J_b,t)=0$ in the $(J_a,J_b)$ plane. They indicate positions of exact CNOT gates plotted as the function of the scaling parameter $0<\lambda \leq 1$. Three coupling constants $(g_{zx},g_{xz},g_{xx})$ are scaled as explained in the text,  while $g_{zz}=1/4$. Trajectories start at $\lambda=0$ at integer $(J_a,J_b)$ values of the same parity and end at $\lambda=1$ at the red (large) circles. Grey (small) circles follow with equal intervals in $\lambda$. Numbers indicate time $t$ at the ends of the trajectories,  represented as $10^3(t/\pi-1)$.}
\label{fig:trajectories}
\end{figure}

\vspace{3mm}

One may notice that in Fig.~2 the zeros of $G$ corresponding to small $J_a=3$ (or $J_b=3$) which are away from the $J_a=J_b$ line, do not generally move towards the diagonal line, as $\lambda$ increases from 0 to 1.  
This suggests that there may be a more complicated behavior for $J_a\leq3$ (and $J_b\leq3$).   In Fig.~3 we illustrate the trajectories for  points originating at the integers (4,2) and (2,4), as $\lambda$ is increased from zero.  In this case the two trajectories meet at the axis $J_a=J_b$, at a common value $J_c $, when $\lambda$ reaches a  critical value  $\lambda_c$  which is smaller  than 1, and we find  no  solutions for  $\lambda > \lambda_c$.   As seen in Fig.~3, the trajectory follows 
a horseshoe-like arc, where at first, both branches of it shift in the direction of small $J_b$ and $J_a$, respectively, but finally merge at $J_a=J_b\equiv J_c$. The inset to the figure shows that the dependence on $\lambda$ of the distance $\delta J$ between the point $(J_a,J_b)$ and the merger point $(J_c,J_c)$ 
can be accurately described by  $\delta J \propto 
 \sqrt{\lambda_c - \lambda}$. 
 
The behavior seen in Fig.~3, as $\lambda$ is varied, is, in fact, qualitatively similar to  the behavior that we found in Sec. \ref{sec:bert}, when $g_{xx}$ was varied, in the exactly solvable case with $g_{xz}=g_{zx}=0$. The dashed curve in Fig.~3 shows the hyperbolic trajectory $J_a J_b = rs = 8$,   obtained in the exactly solvable case when $(r,s)$ = (4,2)

\vspace{3mm}

\begin{figure}[!hbtp]
\includegraphics[width=3.0in]{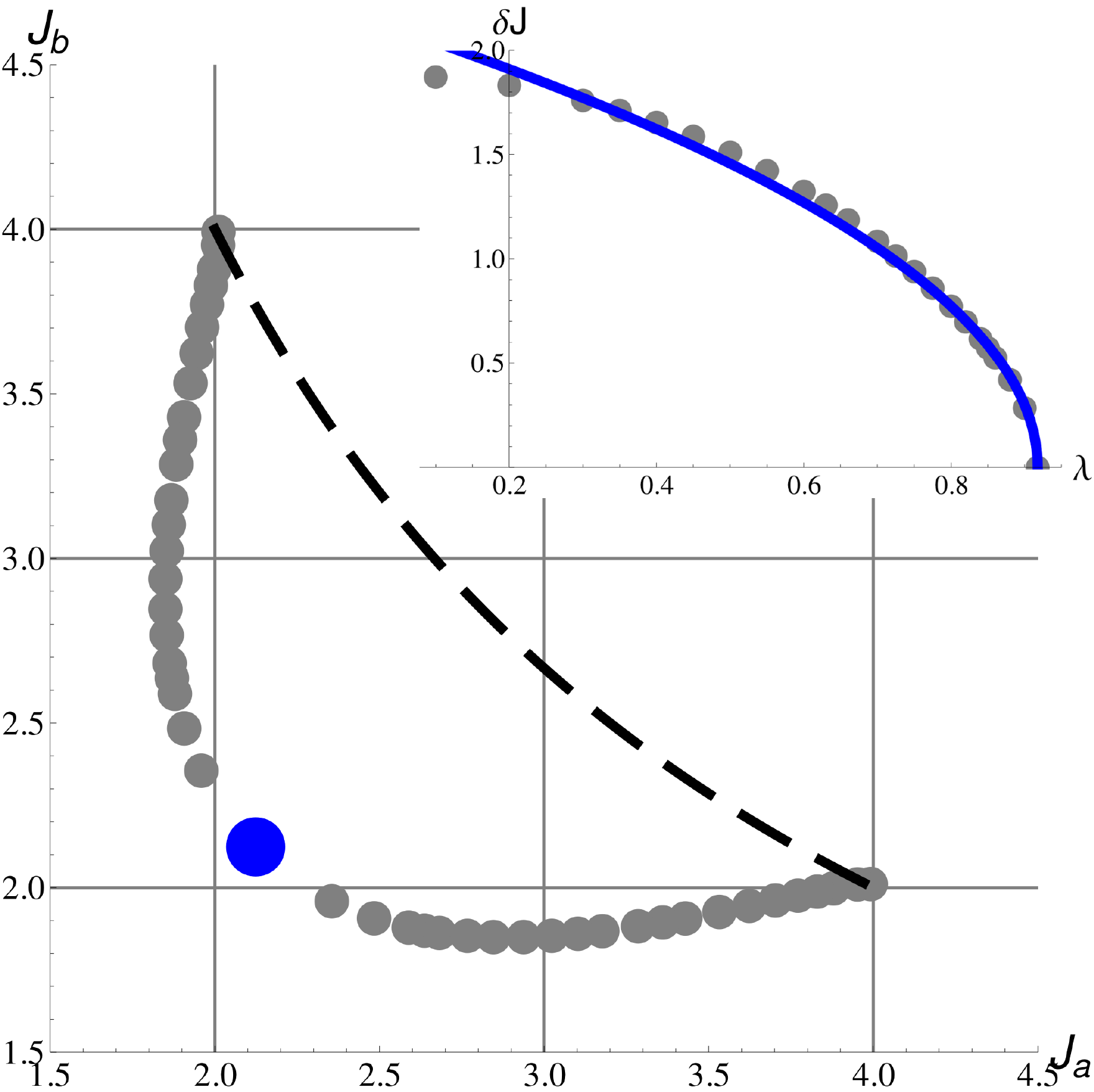}
\caption{(Color online) Grey dots show the trajectory of results for $(J_a,J_b)$, connecting points $(4,2)$ and $(2,4)$, as $\lambda$ is varied.   Ends of the arc correspond to the scaling parameter limit $\lambda\rightarrow0$. Branches of the arc meet at the blue (large) point at the $J_a=J_b$ diagonal at $J_c\approx 2.1239$, $\lambda_c \approx0.9176$, and $t\approx 1.332 \pi$.    Inset: Distance $\delta J$ of $(J_a,J_b)$ from the point 
$(J_c, J_c)$
 as a function of $\lambda$. Near the critical value $\lambda_0$ the data shown by grey (small) dots go as $\delta J\propto\sqrt{\lambda_c-\lambda}$; this dependence is shown by a blue (full) curve. The dashed curve in the main figure shows the trajectory of $(J_a,J_b)$ for the exactly solvable case where $g_{xz}=g_{zx}=0$, when $g_{xx}$ is varied, starting from the same points (4,2) and (2,4).} 
\label{fig:4-2Arc}
\end{figure}

\vspace{3mm}

For the exactly solvable case, with $g_{xz}=g_{zx}=0$, we found that for {\em any} choice of the integers $(r,s) $, there is a critical value of $g_{xx} $, given by $|r-s|/2$,  beyond which  solutions cease to exist.  It is a reasonable conjecture that  similar behaviors will occur if we increase $\lambda$  sufficiently in the model with all couplings nonzero. We have checked this in the case where one starts from $J_a=6, J_b=4$ or vice versa; here, we find that  the trajectory ends at $J_a=J_b \approx 4.33886, t  \approx 1.12248 \pi$, when $\lambda$  reaches a critical value
$\lambda_c \approx 1.15147$.
However, it is possible that the trajectories will behave differently for other starting points.

Of interest is not only the position of zeros of $G$ but also the behavior of functions $G_1$ and $G_2$ near these zeros. In the close vicinity of zeros, the functions are defined by their derivatives. All three first derivatives of $G_1$ at these points vanish. All six second derivatives do not vanish and are real, the derivative $\partial^2G_1/\partial t^2$ is the largest one. Imaginary terms appear only in third derivatives. Because in the quadratic approximation the tensor of second derivatives is a diadic product of the vector $\nabla{\rm Re}[G_1]$ onto itself, two of its eigenvalues vanish and the third is positive. A similar inspection of the properties of  $G_2$ shows that all three first derivatives vanish and second derivatives are finite. Therefore, the behavior of both functions  is analytic near the zeroes of $G$, and we find that zeros are saddle points of $G_2$. As an example, we quote  the data for an ``even-even" zero of $G$, with $J_a\approx15.63, J_b\approx7.91$, calculated for the basic Hamiltonian of Eq.~(\ref{eq3}). The only nonzero eigenvalue of the matrix of second derivatives of $G_1(J_a,J_b,t)$ is 0.254, and eigenvalues of the matrix of second derivatives of $G_2(J_a,J_b,t)$ are $-4.443,-0.336$, and $0.017$. The latter data characterize the shape of $G_2(J_a,J_b,t)$ surface near its saddle point, and we attribute the difference in the magnitude of three eigenvalues to the dominance of $t$-derivatives over the derivatives with respect to  $(J_a,J_b)$, when  $J $ is large.  

The behavior of $G(J_a,J_b,t)$ at a larger scale can be found only numerically. In all cases,  plots of $G$, with $t$ fixed at its value in one of the zeros of $G$, include valleys in the $(J_a,J_b)$-plane running at an  the angle $ \approx \pi/4$  relative to $J_b$ axis, in agreement with the asymptotic analysis of Sec.~\ref{sec:large}, Eq.~(\ref{eq14}). As seen in the example shown in Fig.~4, the sharpest  valley is the one passing through the $G=0$ point.

\vspace{3mm}

\begin{figure}[!hbtp]
\includegraphics[width=3.0in]{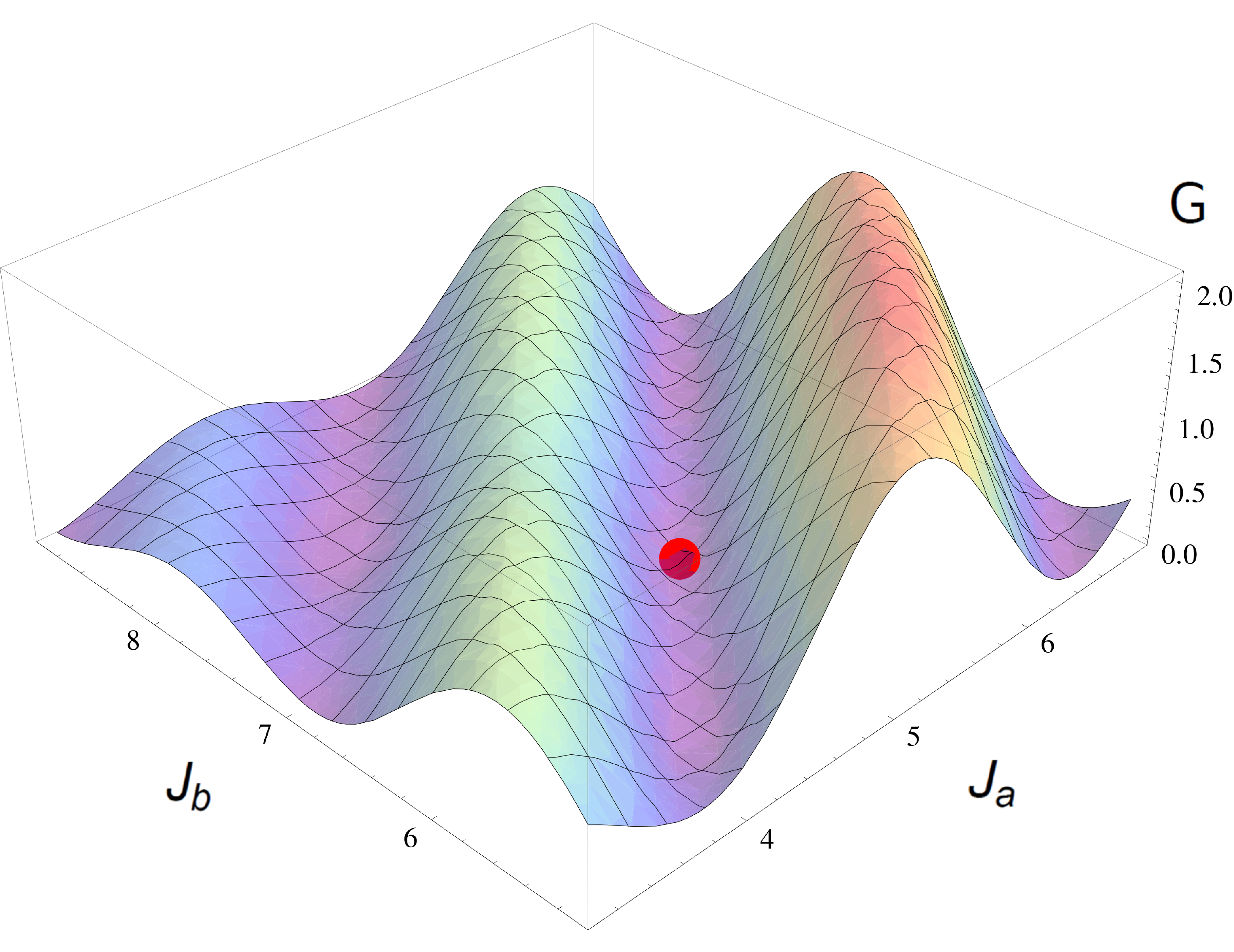}
\caption{(Color online) A plot of $G(J_a,J_b,t)$ as a function of $(J_a,J_b)$ with time $t$ fixed at its value in the zero of $G$ shown by a red (large) point. Parameter values of the $G=0$ point are $J_a\approx5.187, J_b\approx6.564, t/\pi\approx0.993$. The most prominent features of the plot are deep valleys in the direction bisecting the angle between $(J_a,J_b)$ axes.}
\label{fig:valleys}
\end{figure}

\vspace{3mm}

It is seen from Fig.~2 that all zeros of $G$ in the region $J_a,J_b>2.5$ are genetically related to zeros of the asymptotical expansion of Sec.~\ref{sec:large}. This is also true for the arc of Fig.~3. We believe that these zeros will be the ones of greatest  importance for constructing CNOT gates in practice. However, in order  to provide a more complete physical picture, we remark that generically zeros of $G$ exist also in the small $(J_a,J_b)$ region. As an example, we quote here data on zeros of $G$ at the coordinate axes of the $(J_a,J_b)$ plane. For the basic model of Eq.~(\ref{eq3}), the trace of the matrix $m$ can be calculated explicitly and is real, whenever $J_b=0$. As a result, two equations $G_1=0$ reduce to a single equation  
\[\tan{(R_+t/2)}\tan{(R_-t/2)=\frac{R_+R_-}{4-J_a^2}},\,R_\pm=\sqrt{4\pm2Ja+J_a^2}.\]
Remarkably, roots of this equation satisfy the condition $G_2=1$ for the second Makhlin invariant. Because of the periodicity of tangents, solutions of this equation form a multi-branched spectrum $t=t(J_a)$. Similar solutions $t=t(J_b)$ exist for $J_a=0$. We have not carried out an extensive analysis of the solutions that are possible in regions close to one of the axes $J_a=0$ or $J_b=0$, so we must  leave further studies of this problem to a future investigation

\section{$M_d$ and $m$ matrices}
\label{sec:Mdm}

In the previous section, we showed numerically that an infinite set of exact zeroes of $G$ exists for a physical model, capacitively coupled exchange-only qubits. At such zeroes, we can apply the analytic results of Sec~\ref{sec:KC}, including Eq. (\ref{eq19}) and everything related to it.  
In the current section we return to analytical procedures and establish additional results, which are exact at the zeros of $G$ and specific to them.

The matrix $m=M_B^TM_B$ is symmetric by construction and its eigenvectors $|\Psi_j\rangle$ can be chosen {\it real in magic basis} (except of global phases) and are {\it maximally entangled}.\cite{Makhlin} We designate its four-component eigenvectors as $\Psi_j^B$ and choose them to be real. For $1/J\ll1$, off-diagonal elements of matrix $m$ are small, and we numerate $\Psi_j^B$ in such a way that the eigenvector with dominating $j$-component is assigned as $\Psi_j^B$. We then find that in the limit of large $J$, at the zeroes of $G$,  the eigenvalues $\mu_j$ of matrix $m$, $m|\Psi_j\rangle=\mu_j|\Psi_j\rangle$, are 
\be
\mu_1=e^{-i\pi/2},\,\mu_2=e^{i\pi/2},\,\mu_3=e^{i\pi/2},\,\mu_4=e^{-i\pi/2}.
\label{eq20}
\ee
These results remain exact even for finite values of $1/J$, since, as  we have already seen in  
Sec.~\ref{sec:KC}, the eigenvalues of $M_d$ are precisely equal to $e^{\pm i \pi / 4}$ at the zeros of $G$.

After transforming the unitary matrix $M_d$ of Eq.~(\ref{eq15}) to magic Bell basis it turns into a diagonal matrix with eigenvalues $e^{-i\lambda_j}$, $M_d=\sum_{j=1}^4|\Phi_j\rangle e^{-i\lambda_j}\langle\Phi_j|$.\cite{KC2001} Therefore, $M_S$ matrix of Eq.~(\ref{eq15}) in magic basis reads $M_B=U_{ab}^BM_dV_{ab}^B$, where $U_{ab}^B$ and $V_{ab}^B$ are $U_{ab}$ and $V_{ab}$ matrices in magic Bell basis. Then
\be
m=(V_{ab}^B)^TM_d(U_{ab}^B)^TU_{ab}^BM_dV_{ab}^B.
\label{eq21}
\ee
Because matrices $(U_a,U_b)$ and $(V_a,V_b)$ of local rotations are unitary and real,\cite{Makhlin} the same is true for $(U_{ab},V_{ab})$ matrices, hence
\be
m=(V_{ab}^B)^{-1}M_d^2V_{ab}^B.
\label{eq22}
\ee
Therefore, the matrices $m$ and $M_d^2$ are unitary equivalent and their spectra coincide.\cite{FNER2}  Writing $\mu_j$ of Eq.~(\ref{eq20}) as $\mu_j=e^{2i\epsilon_j}$, with $-\pi/2\leq\epsilon_j\leq\pi/2$, and eigenvalues of $M_d$ found at the end of Sec.~\ref{sec:KC} as $e^{-i\lambda_j}$, we arrive at $e^{2i(\lambda_j+\epsilon_j)}=1$. Extracting the square root of this equation as $e^{i(\lambda_j+\epsilon_j)}=1$ (because $|\epsilon_j+\lambda_j|<\pi$), we arrive at equations
\be
\epsilon_j+\lambda_j=0
\label{eq23}
\ee
relating all $\lambda_j$ to $\mu_j$ of Eq.~(\ref{eq20}). Finally, we arrive at an exact expression for $M_d$ matrix valid at all zeros of $G$
\be
M_d=
\left(\begin{array}{cccc}
e^{-i\pi/4}&0&0&0\\
0&e^{i\pi/4}&0&0\\
0&0&e^{i\pi/4}&0\\
0&0&0&e^{-i\pi/4}
\end{array}\right)
\label{eq24}
\ee
with ${\rm Tr}[M_d]=2\sqrt{2}$ in agreement with Sec.~\ref{sec:KC}. We note that while absolute values of phases $|\lambda_j|=\pi/4$ were known from Sec.~\ref{sec:KC}, Eq.~(\ref{eq24}) ascribes proper signs of phases to all matrix elements of $M_d$ in the magic basis.

 Remarkably, the Eq.~(\ref{eq24}) for $M_d$ is valid both in the magic and standard bases because matrix elements of $M_d$ obey relations $M^d_{11}=M^d_{44}$ and $M^d_{22}=M^d_{33}$. This universal form of $M_d$ in the standard basis is important for calculating local rotations, see Sec.~\ref{sec:LR} below.

In the solvable  case of Section \ref{sec:bert},  where $g_{xz} =g_{zx} = 0$, but $g_{xx} \neq 0$, we found that exact CNOT gates are possible at a discrete set of points, with $t =\pi$, and values of $J_a,J_b$ given by  Eqs. (\ref{eqEE}).
 At these points, the matrix $M_S$ has a simple diagonal form, with diagonal elements 
$$
M^S_{11} = M^S_{44} = (-1)^{(r+s)/2}  \,  e^{-i \pi /4} ,
$$
\be
\label{MSB}
M^S_{22} = M^S_{33} = (-1)^{(r-s)/2}  \,  e^{i \pi /4} .
\ee
In this case, the eigenvectors $|\Psi_j\rangle$ can be chosen equal  to $|\Phi_j\rangle$, and $M_B=M_S$.  However, $M_d$ and $M_B$   differ from each other by  local rotations, except in the case where $s$ is even and $r-s$ is a multiple of four.

In conclusion, we find that  at all the zeroes of $G$, the matrices $M_d$ and $m$ can be put in universal diagonal forms, with eigenvalues defined by 
Eqs.~(\ref{eq24}) and (\ref{eq20}), respectively.

\section{Local rotations}
\label{sec:LR}

It has been shown above that the matrix $M_S$ becomes equivalent to an exact CNOT gate, up to local rotations, at  an infinite set of points $(J_a,J_b,t)$. In this section, we dicsuss the explicit form of the necessary local rotations, in some simple cases.

Entanglement originates from the Hamiltonian $H_{\rm int}$ of Eq.~(\ref{eq2}), which, in general,  includes both diagonal and off-diagonal parts. In the special case of the butterfly geometry, the Hamiltonian $H_{\rm int}$ reduces to a diagonal Ising coupling,  $H_{\rm diag}=g_{zz}\sigma_z^a\sigma_z^b$, which can  provide a CNOT gate for $t=\pi$ at arbitrary values of $(J_a,J_b)$. By contrast, as we  proved above,  adding to  $H_{\rm int}$ an off-diagonal part, which will be  present in generic geometries, restricts exact CNOT gates to a countable set of the values of parameters $(J_a,J_b,t)$. 

Because $M_S(t)=\exp{(-iH_{\rm diag}t)}$ is diagonal in the standard basis, it is easy to find local rotations accompanying the nonlocal part, in the Ising case.  Because the Controlled-Z matrix is also diagonal in the standard basis, with the diagonal of $(1,1,1,-1)$, the local part relating them includes only $z$-rotations. If we choose $z$-rotations of individual qubits in the standard form as $U^z_{a,b}=\left(\begin{array}{cc}e^{i\phi_{a,b}/2}&0\\0&e^{-i\phi_{a,b}/2}\end{array}\right)$, then $U_{ab}^z=U_a^z\otimes U^z_b$. Next, with $t=\pi$, the product $M_SU_{ab}^z$ is equal to the Controlled-Z matrix, up to a global phase, when $\phi_a=\frac{\pi}{2}(1-2J_a)$ and $\phi_b=\frac{\pi}{2}(1-2J_b)$. Choosing $(J_a,J_b)$ as integers of the same parity, as in Sec.~\ref{sec:large}, we find equal $\pi/2$ rotations of both qubits, $\phi_a=\phi_b=\pm\pi/2$, with the rotation direction depending on the parity of $(J_a,J_b)$. More generally, the magnitudes of the rotation angles are controlled by the diagonal part of $H_{\rm int}$, chosen as $\frac{1}{4}\sigma_z^a\sigma_z^b$ as everywhere above, and by the choice of $(J_a,J_b)$.

In the more general solvable  case of Section \ref{sec:bert},  where $g_{xz} =g_{zx} = 0$, but $g_{xx} \neq 0$, we found that exact CNOT gates are possible at points with $t =\pi$ and values of $J_a,J_b$ given by  Eqs.~(\ref{eqEE}). At these points, the  matrix $M_S$ has the diagonal form given by  (\ref{MSB}), and we find, once  again,   that we can choose the local rotations to be rotations about the z-axis, with $\phi_a = \phi_b = \pm \pi /2$, depending on the parity of the integer $s$.

When $g_{xz}$ and $g_{zx}$ are nonzero, 
 exact CNOT gates are still possible at an infinite set of points $(J_a,J_b,t)$ as was shown in Sec.~\ref{sec:CNOT}. However, under these conditions $M_S$ will no longer be diagonal.   For large $J$, the off-diagonal elements will be of order $1/J$, just as we found that  $J_a$ and $J_b$ deviate from integer values, and $t$ deviates from $\pi$, by amounts of order $1/J^2$.  
 Similarly, the local rotations $U_{ab}$, and $V_{ab}$  will  acquire corrections of the order of $1/J$ that should be found for each zero of $G$ by the Kraus-Cirac procedure.\cite{KC2001} With $M_S$, $U_{ab}$, and $V_{ab}$ calculated for some zero of $G$, Eq.~(\ref{eq15}) brings us to $M_d$ of Eq.~(\ref{eq24}), which can be easily transformed to the Controlled-Z form as described in the previous paragraphs. Because the small rotations arising from the off-diagonal part of $M_S$ are specific for each $(J_a,J_b,t)$ zero of $G$, we do not calculate them here.

\section{Summary}
\label{sec:summary}

The theory of quantum circuits includes two types of operations, local and nonlocal. At the heart of it are nonlocal two-qubit operations because they produce quantum entanglement. Local operations play an auxiliary role. The CNOT gate is a paradigmatic nonlocal gate because it is universal and allows performing arbitrary rotations in multiqubit systems.  The controlled-Z gate is equivalent to CNOT, from this point of view,  because the two gates  differ from each other  only by local rotations   (two Hadamard  rotations). Because nonlocal operations critically depend on interqubit coupling, which is relatively weak in many realizations, the efficiency of these operations may be  a limiting factor in the speed of quantum computation. Thus, it is helpful to see how to most efficiently employ the available interqubit couplings to perform any  desired two-qubit operation.

An important example is the exchange-only coded qubit, where  information is encoded in a  two-dimensional  subspace with $S=S_z =1/2$,  within the  eight-dimensional   low-energy  Hilbert space for the three participating electron spins. Since single qubit operations are implemented by exchange coupling in these systems,  they can be performed quite rapidly.  If two-qubit coupling is also implemented by exchange, however,  the spin quantum numbers of individual qubits will not be conserved, and the system will not generally remain in the coding Hilbert space.  Then, to rectify this, one is forced to implement a complicated series of at least 15 non-local operations, interspersed with local operations,\cite{Setiawan,Mehl} all of  which take time and can easily lead to errors.   By contrast, if the interqubit couplng is  mediated only  by the Coulomb interaction between electrons in different qubits, the spin quantum numbers of each qubit are conserved  by this interaction.  Therefore, if one can neglect perturbations such as spin-orbit coupling and hyperfine coupling to nuclear spins, the qubits will remain in their  coding spaces, and complicated remedial steps will not be necessary.   Thus, it may be highly advantageous to use capacitive coupling based on the Coulomb interactions, even though it is weaker than exchange coupling and the  coupling time for a single step  will be longer.   

In this paper we have explored the issue of whether it is possible, in general, to execute a perfect CNOT (or controlled-Z)  gate with a single time interval of capacitive coupling between the two qubits. In general, there will be four parameters , denoted by $g_{zz}, g_{xx}, g_{zx}, g_{xz}$, which describe this capacitive coupling, whose values may be fixed by geometrical constraints, and may be difficult to vary independently. Here we have assumed that the couplings may be turned on and off  sharply but that their on-values, and the ratios between them,  are fixed by the geometry. We have found that for a wide range of the coupling parameters,  a perfect CNOT  gate is indeed possible, provided we choose appropriately the single qubit exchange splittings $J_a$ and $J_b$ as well as the time $t$ that the coupling is turned on. In fact, we have found in each case  that there is a countably infinite set of triplets $(J_a, J_b, t)$ that satisfy the criteria.  We have also discussed, in some detail, how the precise values of  $(J_a, J_b, t)$ vary as one varies the assumed ratio between the four  capacitive coupling constants, and in special cases,  we have found  the local rotations necessary to complete a controlled-Z operation.

Although we have focused on exchange-only spin qubits, other applications may also be considered. 
Mathematically, our considerations should apply to any system of qubits where the coupling can be written in the form of Eq.~(\ref{eq2}), to a good approximation, and there is a single qubit splitting of the form of Eq.~(\ref{eq1}).  

As a specific example, one may consider a system of $S-T_0$ qubits, where
the coding qubit states are linear  combinations of  a singlet state and a triplet state with $S_z=0$, for two electrons in a double quantum dot.    In order to perform the full array of single qubit operations, it is necessary to create a local magnetic field gradient, so that electrons in the two dots see different magnetic fields.  If the magnetic field on both dots is in the same direction, and if we can neglect effects such as spin-orbit coupling and transverse components of the nuclear Overhauser field, then capacitive coupling between two  qubits will conserve $S_z$ for each qubit separately, so the system will remain inside the coding subspace. Thus there would be a  possibility to perform an exact  CNOT gate in this system in a similar fashion to the case of exchange-only qubits.  Calderon-Vargas and Kestner have in fact considered this possibility in a recent work,\cite{calderon} and have shown numerically,  that  exact CNOT gates can be achieved with a single interval of non-local coupling by adjusting three parameters, the coupling time and the voltage bias on each of the two qubits, in several examples.

The mathematical examples considered in this paper have assumed that the interqubit coupling constants can be turned on and off instantaneously.  A more realistic scenario would be a case where the coupling is turned on and off over a  time interval $\tau>0 $ that is short compared to the coupling duration $t$.    According to our understanding of the analytic structure of the problem, this should not affect our basic results. Specifically, if $\tau$ is  sufficiently small, there should still exist triplets $(J_a, J_b, t)$ for which the Makhlin invariants {\em exactly} satisfy the criteria for an exact CNOT gate, though the values will be displaced from the solutions for $\tau=0$.  If $\tau$ becomes too large, however, pairs of solutions may come together and disappear.  We conjecture that a necessary condition for the existence of exact solutions might be that $J_a \tau$ and $J_b\tau$ be small compared to unity.  Calculations presented in Ref.~\onlinecite{calderon} have indeed shown the possiblity of achieving  exact CNOT gates in a protocol with a finite time $\tau$ for turning on and off the control voltages, 
in the case of $S-T_0$ qubits.

An assumption, in all of our models, is that we could confine ourselves to a low energy-Hilbert space with two states for each qubit.  This assumption can be justified provided there is a large energy gap $E_g$ to all other states with the same conserved quantum numbers as the desired qubit states, and provided variations of the coupling constants are always adiabatic compared to  $1 /E_g$.  In the present case, this means we must have $E_g \gg t^{-1} > J$ .   Although, generically,  the conditions for a CNOT gate will  no longer be satisfied exactly when higher energy states are taken into account, we expect that  errors can be made  arbitrarily  small, for large enough $E_g$, as errors due to non-adiabaticity should fall off faster than any power of  $1/(t E_g)$.

The possibility to perform CNOT gates of arbitrary perfection using a single interval of  capacitive coupling, giving  a huge decrease in the number of  nonlocal operations relative to the case of exchange coupling,  sounds highly promising.  This calls for new efforts in enhancing and better control of capacitive interqubit coupling. Floating electrostatic gates\cite{static} and circuit quantum electrodynamics\cite{CQE} are among the currently discussed tools.

\section{Acknowledgments}

Research was supported by the Office of the Director of
National Intelligence, Intelligence Advanced Research
Projects Activity (IARPA), through the Army Research
Office Grant No. W911NF-12-1-0354 and by the NSF
through the through the Materials World Network Program
No. DMR-0908070.  We acknowledge helpful discussions with C.~M.~Marcus, J.~Medford, J.~M.~Taylor, J.~I.~Cirac, and, especially, Barry Mazur, and we  thank F. A. Calderon-Vargas for bringing  Ref. \onlinecite{calderon} to our attention.


\begin{thebibliography}{99}
\bibitem{footnote} The model of two coupled Josephson qubits solved in Ref.~\onlinecite{Makhlin} and \onlinecite{ZVSW} can be obtained from the Hamiltonian of Sec.~\ref{sec:bert} by putting $J_a=J_b$ and $g_{zz}=0$. Its solutions for CNOT gates depend on two discrete quantum numbers.
\bibitem{FNER} As distinct from the invariants $(G_1,G_2)$ that are periodic in all $\alpha_j$ with a period $\pi/2$, $\exp{(-iH_d)}$ has a longer period of $2\pi$. This suggests that $M_d$ is defined only up  to phase factors that can be included into $(U_{ab},V_{ab})$ matrices. Our definition of $M_d$ here and in what follows corresponds to the choice of $\alpha_j$ according to Eq.~(24).
\bibitem{FNER2} While the spectra of $m$ and $M_d$ are connected by a simple relation,  $\mu_j=\exp{(-2i\lambda_j)}$, we emphasize that the matrices are diagonal in different  bases:   $m$ in the $|\Psi_j\rangle$ basis and $M_d$ in the magic basis $|\Phi_j\rangle$.
\end{thebibliography}
\end{document}